\documentclass{article}

\usepackage{graphicx}
\usepackage{caption}
\usepackage{subcaption}
\usepackage{adjustbox}

\usepackage{PRIMEarxiv}
\usepackage{amsmath}
\usepackage[utf8]{inputenc} 
\usepackage[T1]{fontenc}    
\usepackage{hyperref}       
\usepackage{url}            
\usepackage{booktabs}       
\usepackage{amsfonts}       
\usepackage{nicefrac}       
\usepackage{microtype}      
\usepackage{lipsum}
\usepackage{fancyhdr}       
\usepackage{graphicx}       
\graphicspath{{media/}}     

\title{Predicting Breast Cancer Phenotypes from Single-cell RNA-seq Data Using CloudPred
}

\author{
  Hossein Moghimianavval \\
  Department of Mechanical Engineering \\
  University of Michigan \\
  Ann Arbor\\
  \texttt{mhossein@umich.edu} \\
   \And
   Baharan Meghdadi \\
  Department of Chemical Engineering \\
  University of Michigan \\
  Ann Arbor\\
  \texttt{baharan@umich.edu} \\
   \And
   Tasmine Clement \\
  Department of Computational Medicine and Bioinformatics \\
  University of Michigan \\
  Ann Arbor\\
  \texttt{tasmine@umich.edu} \\
   \And
   Man I Wu \\
  Robotics Department \\
  University of Michigan \\
  Ann Arbor\\
  \texttt{maniwu@umich.edu} \\
   \And
}

\begin{document}
\maketitle

\begin{abstract}
\textbf{Motivation:} Numerous tools have been recently developed to  predict disease phenotypes using single-cell RNA sequencing (RNA-seq) data. CloudPred is an end-to-end differentiable learning algorithm coupled with a biologically informed mixture model, originally tested on lupus data. This study extends CloudPred's applications to breast cancer disease phenotype prediction to test its robustness and applicability on untested and unrelated biological data. \\
\textbf{Results:} When applying a breast cancer single-cell RNA seq dataset, CloudPred achieved an area under the ROC curve (AUC) of 1 in predicting cancer status and performed better than a linear and Deepset model.
\end{abstract}


\section{Introduction}
Over the past few decades, single-cell RNA sequencing (scRNA-seq) tools have been able to better inform us about gene expression at the individual cell level which has, in turn, provided us a deeper understanding of disease prognosis\cite{palSinglecellRNAExpression2021}.

Given that scRNA-seq tools have the potential to make powerful inferences regarding individual health in research and clinical settings, it is important that the developing machine learning algorithms are robust, interpretable, and generalizable across different data collections. A major setback of common scRNA-seq analysis pipelines is that they require same number of cells from each patient which is often not possible. Consequently, synthetic data generation steps are commonly implemented in such analysis models to address such disparities. Here, we utilize a machine learning model for analyzing scRNA-seq data called CloudPred\cite{he2021arxiv}. CloudPred analyzes each patient scRNA-seq data using a Gaussian mixture model and extracts certain subtypes with varying prevalence that is independent of patient's scRNA-seq total number of cells. CloudPred then utilizes the subtype prevalence for each patient to predicts their state (healthy or diseased). CloudPred initially highlighted a real scRNA-seq dataset of 142 lupus patients and controls to test the capabilities of the model. The prediction capabilities of CloudPred were also compared to several other alternative prediction tools: a linear model, Deepset, a permutaton-invariant method. 

Although in the original work\cite{he2021arxiv}, CloudPred showed good performance against several other methods, it is unclear whether the model, initially trained for lupus, is useful for predicting other diseases. In order to transition machine learning models from research settings into the clinical setting, models must show that that they are generalizable before being implemented in clinical settings. Therefore, a more robust practice to provide support for machine learning prediction models is to test a model with different datasets. In this way one can get data with as many diverse patient samples as possible to train a deep learning model that works across multiple clinical situations. While this seems like a formidable goal for most researchers, it is important to start off by capturing a model's generalizability by applying models to multiple diseases after they are presented in studies that only feature one disease. 

Here, we investigated the performance and generalizability of CloudPred, which was developed using a lupus scRNA-seq dataset, on a breast cancer dataset representing a new biological dataset. Similar to lupus dataset, breast cancer dataset included scRNA-seq data from patients that contained normal/healthy and diseased phenotypes. We also compared the ability to predict breast cancer phenotypes from sc-RNA seq using the alternative methods of linear prediction and Deepset, a permutaton-invariant method. 

\section{Approach}\label{sec2}

The approach was to analyze an unrelated scRNA-seq data collection using CloudPred. This new dataset provides data on samples from triple negative breast cancer (TNBC) patients from a separate study in 2021 in which the primary focus was to examine gene expression of normal and tumorigenic cells \cite{palSinglecellRNAExpression2021}. Although many scRNA-seq studies have focused on the heterogeneity of tumor tissues or differences between the paired patients’ tumor and normal tissues, the availability of scRNA-seq from healthy donors is limited. As a result, we have utilized data from 21 patients (8 with TNBC and 13 normal) from this study in order to test the generalizability of CloudPred. We used this data to train CloudPred with 11 samples and validate it with 5 samples. 5 samples (3 healthy, 2 with TBNC) were kept unseen for the final testing. We also compared our results with both a linear and Deepset models. 

\section{Methods}\label{sec3}

\subsection{Data Preprocessing}\label{subsec2}

To identify cell types, we followed the data preprocessing steps from \cite{chenCodeDownstreamAnalysis2022}. We used filtered counts comprised of 54,332 and 54,820 cells from healthy donors and TNBC patients, respectively. Filtered gene expression matrices were merged for normal and tumor samples, separately, and then log normalized and scaled. The \texttt{FindVariableFeatures} in Seurat was used to select the first 2000 highly variable genes. The anchor-based Seurat integration was used to remove batch effects and find common cell types among samples (functions \texttt{FindIntegrationAnchors} and \texttt{IntegrateData} with default parameters). This was followed by \texttt{FindNeighbours, FindClusters}, and \texttt{RunTSNE} functions in Seurat with default parameters. The cell annotations were done in two, and three levels for normal, and tumor samples, respectively. 

\begin{figure*}
    \centering
    \includegraphics[]{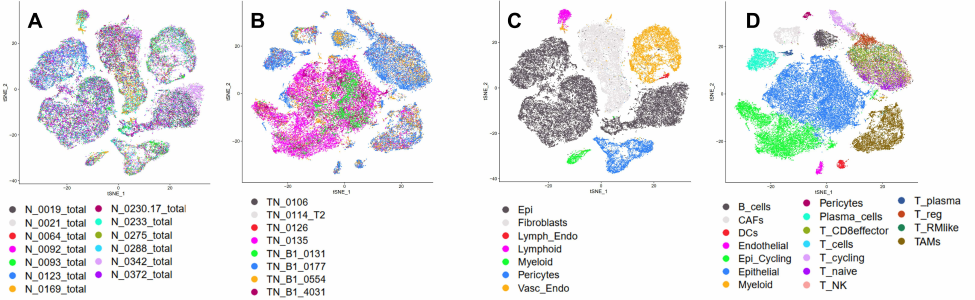}
    \caption{Visualization of single cell breast cancer atlas. (A) t-SNE plot shows the integration of 13 scRNA-seq profiles of healthy donors. (B) t-SNE plot shows the integration of 8 scRNA-seq profiles of TNBC patients. (C) t-SNE plot of normal samples colored by cell types. (D) t-SNE plot of tumor samples colored by cell types.}
    \label{fig:tsne_normal_tnbc}
\end{figure*}

First, the total cells were clustered with a low resolution to annotate epithelial and non-epithelial cells, given their marker genes. For tumor samples, the epithelial cells with higher expression of MKI67,a proliferation marker, were annotated as cycling epithelial cells. The non-epithelial cluster was subset and clustered again to annotate immune and stromal cell lineages in more detail which include fibroblasts, vascular-endothelial cells, pericytes, myeloids, and lymphoids in normal samples, and cancer-associated fibroblasts (CAFs), pericytes, endothelial cells, dendritic cells (DCs), plasma cells, B cells, T cells, and tumor-associated macrophages (TAMs) in tumor samples. The T cell population in tumor samples was subset and clustered again to annotate natural killer cells (NK), CD8+effector, naïve, regulatory, cycling, and tissue-resident T cells. The cell type markers can be found in \cite{palSinglecellRNAExpression2021}. 

CloudPred requires the gene expression matrix with the size of number of genes $\times$ number of cells and a vector of cell types with the size of number of cells. The cell annotations were saved in npy format along with the raw counts of 2000 highly variable genes in npz format, for each patient separately. The normal and cancer patient data was organized in two folders for CloudPred analysis. The raw counts were further transformed using CloudPred's built-in principal component analysis that reduces the dimension of input data for downstream analysis to $n\times p$ where $n$ is the number of cells in each patient sample and $p$ is the number of principal components (defaults to 100).   


\subsection{CloudPred workflow}

Assuming that the patient's scRNA-seq data follows a multivariate Gaussian mixture distribution, CloudPred attempts to find the parameters of each distribution, namely mean ($\mu_k$) and variance ($\Sigma_k$), and their corresponding weights,$w_k$, for a predefined $K$ number of Gaussian distributions or components. For simplicity, CloudPred assumes there is no correlation between variables, making the covariance matrix of each component diagonal. Practically, this assumption means that every patient's scRNA-seq data is composed of $K$ subtypes, each consisting different cell types. Therefore, the probability of a single cell $\mathbf{x}$, belonging to component $k$ follows a normal distribution:
\begin{equation*}
    Pr(\mathbf{x}|\mathbf{\mu_k},\mathbf{\Sigma_k}) \sim \mathcal{N}(\mathbf{\mu_k},\mathbf{\Sigma_k})
\end{equation*}
The model then calculates the log-likelihood of each component for every cell utilizing the log-likelihood function for multivariate Gaussian distributions:
\begin{equation*}
    Pr_k(\mathbf{\mu_k},\mathbf{\Sigma_k};\mathbf{x}) = -\frac{1}{2}[\log(2\pi) + \sum_i^p \log(\sigma_i^2) + \sum_i^p\log(\frac{x_i-\mu_i}{\sigma^2_i})]
\end{equation*}
Where $\mathbf{\mu_k}$ and $\mathbf{\Sigma_k}$ stand for a $p-$dimensional mean vector and a $p\times p$ diagonal variance matrix for component $k$. The $K$ different probabilities are then normalized for each cell with the corresponding weight for each component:
\begin{equation*}
    P_{i,k} = \frac{w_k Pr_k(\mathbf{\mu_k},\mathbf{\Sigma_k};\mathbf{x_i})}{\sum_j^K w_j Pr_j(\mathbf{\mu_j},\mathbf{\Sigma_j};\mathbf{x_i})}
\end{equation*}
The resulting vector $\mathbf{P_i}$ for a single cell $\mathbf{x_i}$ represents the probability of that cell belonging to different subtypes. These probabilities are then averaged over all cells for a patient to generate a vector $s$ that represents the subtype distribution for the patient:
\begin{equation*}
    s = \frac{\sum_i^n \mathbf{P_i}}{n}
\end{equation*}
Following this approach, each patient, regardless of their scRNA-seq total number of cells, will have a $K-$dimensional feature vector which can be used for classification purposes. CloudPred uses a 2nd order polynomial function called $f_\theta$ to generate a logit value for the patient which is next used for predicting the patient's state (diseased or healthy). 
\begin{gather*}
    f_\theta = \sum_i^K a_i s_i + \sum_i^K b_i s_i^2 + c_i\\
    P(f_\theta) = \frac{1}{1+e^{f_\theta}}
\end{gather*}

Cross-entropy loss is then used to calculate the loss between CloudPred predictions and true state. The 2nd order polynomial has $2K +1$ parameters corresponding to the bias value and coefficients of each subtype for each polynomial order. These parameters along with Gaussian mixture parameters ($K(2p+1)$ parameters) are learned during training. During the training process, CloudPred begins with an unsupervised Gaussian mixture model, but then adjusts the parameters of the components to improve the prediction accuracy. Because all parameters can be optimized using differentiation approaches, common machine learning tools such as PyTorch toolbox can be used to implement CloudPred. 
\begin{figure*}[btp]
\captionsetup{justification=centering}
    \centering
    \hspace{-20mm}
    \begin{subfigure}[b]{0.4\textwidth}
        \centering
        \includegraphics[width=0.85\linewidth,height = 0.8 \linewidth]{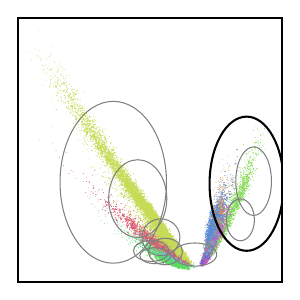} 
        \caption{PCA} \label{fig:CPloss}
    \end{subfigure}
    \hspace{-15mm}
    \begin{subfigure}[b]{0.4\textwidth}
        \centering
        \includegraphics[width=0.85\linewidth,height = 0.8 \linewidth]{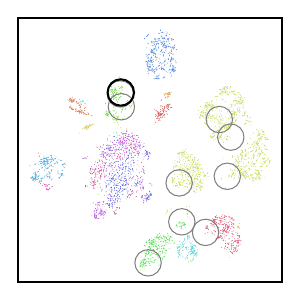} 
        \caption{tSNE} \label{fig:DSloss}
    \end{subfigure}
    \hspace{-8mm}
    \begin{subfigure}[b]{0.25\textwidth}
        \centering
        \includegraphics[width=1\linewidth,height = 1.15\linewidth]{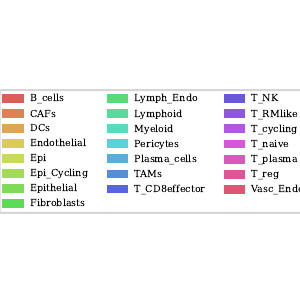} 
    \end{subfigure}
    \caption{Breast cancer dataset projected on two first principle components and a 2-component tSNE. Data from all test patients is visualized. Each ellipse represent a component of Gaussian mixture model. The highlighted ellipse indicates the highly predictive component.  \label{fig2}}
    \label{fig:pca_tsne}
\end{figure*}

\subsection{Benchmarking CloudPred performance for breast cancer scRNA-seq dataset}
Using sklearn python library, CloudPred reports the area under the receiver operating characteristics (AUROC). Along with CloudPred, we report the AUC of two alternative prediction methods: linear and Deepset.

The linear model is an alternative methods that makes the assumption that cells from a patient are independent from each other. As a result, the predictions are made separately for each cell and then the cell-wise predictions are averaged together. The predictions for the linear model are given by the following equation: 
\begin{equation*}
    \mathbf{P_i} = \frac{1}{n} \sum_{i = 1}^{n} w^Tx_i
\end{equation*}
where $\mathbf{w}$ represents the model parameters that are shared across all cells, $\mathbf{w^T}$ represents the transpose of the weight vector $\mathbf{w}$, $\mathbf{x_i}$ represents the input features for the i-th cell, $\mathbf{w^Tx_i}$ is the dot product of the weight vector $\mathbf{w}$ and the input features $\mathbf{x_i}$ for the i-th cell, $\mathbf{\sum_{i = 1}^{n} w^Tx_i}$ is the sum of all dot products over all cells, and $\mathbf{\frac{1}{n}}$ is the scaling factor which divides the sum of the total number of cells $\mathbf{{n}}$ providing an average prediction. 

Deepset is another alternative deep learning model that is the most complex of the three methods. Deepset has a diverse set of applications for supervised, unsupervised, and permutation-invariant tasks. For our purposes, we implemented the permutation-invariant function of Deepset which is able to handle permutations of data within sets. Permutation invariance within Deepset is an important feature because the order of elements in a set should not affect our binary classification of determining cancer phenotypes. For our purposes, each set corresponds to the scRNA-seq data from one patient and the information from individual cells in a patient's scRNA-seq data is treated as a collective set. Some limitations regarding Deepset is that it requires a large and diverse training set for the technique to adequately learn and the results can be difficult to interpret. 

\section{Results}\label{sec3}
We replicated the scRNA-seq analysis of the scRNA breast cancer atlas \cite{palSinglecellRNAExpression2021} to store the gene expression matrices and cell type annotations which are required for CloudPred inputs. To assure the data integration has performed correctly, we plotted normal and tumor samples colored by patient IDs in \textbf{Figs.~\ref{fig:tsne_normal_tnbc}A} and \textbf{B}. Each Seurat cluster contains cells from different patients which confirms the removal of batch effects. Comparing Fig.~\ref{fig:tsne_normal_tnbc}A with Fig.~\ref{fig:tsne_normal_tnbc}B, various batches spread out across the t-SNE plot colored by normal samples which while in t-SNE plot colored by tumor samples, some batches are in a closer proximity. This behaviour elucidates the unique heterogeneity and characteristics of one patient's cancer compared to the others', in contrast to the normal samples which has a closer expression profiles.
Figs.~\ref{fig:tsne_normal_tnbc}C and D show the cell annotations for normal and tumor samples, respectively, which are used in CloudPred plots.

We applied three different models to our breast cancer scRNA-seq dataset and compared the performance of CloudPred with the other two models: a simple linear model and Deepset. All three models performed significantly well in predicting patient states on test dataset and had an AUC of $1$ (\textbf{Table} \ref{tab1}). CloudPred achieved the highest accuracy of 1.0, while the Deepset and linear models had accuracies of 0.94 and 0.95, respectively.

\begin{table}[!t]
\caption{Comparison of model performance on breast cancer dataset \label{tab1}}%
\begin{tabular*}{\columnwidth}{@{\extracolsep\fill}lccl@{\extracolsep\fill}}
\toprule
Method &  AUC & Accuracy \\
\midrule
CloudPred    & 1.0 &   1.0 \\
Deepset    & 1.0   &  0.94 \\
Linear    & 1.0    &  0.95\\

\end{tabular*}

\end{table}

\begin{figure}[htp]
    \centering
    \begin{adjustbox}{minipage=\linewidth,scale=1}
    \begin{subfigure}[t]{0.3\textwidth}
        \centering
        \includegraphics[width = \linewidth,height = \linewidth]{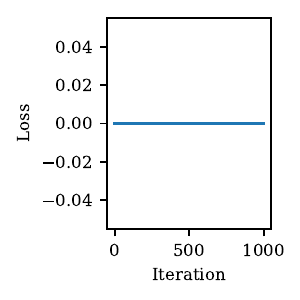} 
    \end{subfigure}
    \hfill
    \begin{subfigure}[t]{0.3\textwidth}
        \centering
        \includegraphics[width=\linewidth,height =\linewidth]{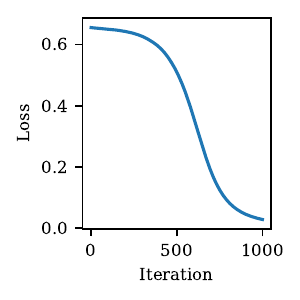} 
    \end{subfigure}
    \hfill
    \begin{subfigure}[t]{0.3\textwidth}
        \centering
        \includegraphics[width=\linewidth,height =\linewidth]{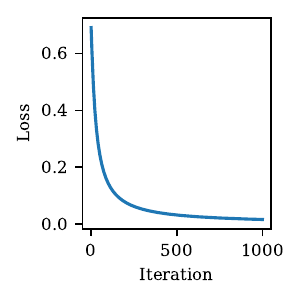} 
    \end{subfigure}
    \vspace{0.5cm}
    \begin{subfigure}[t]{0.3\textwidth}
        \centering
        \includegraphics[width=\linewidth,height = \linewidth]{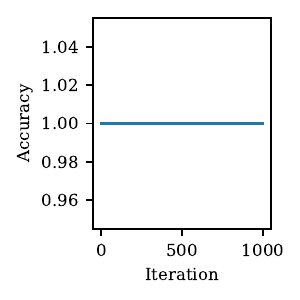} 
        \caption{CloudPred} \label{fig:CPloss}
    \end{subfigure}
    \hfill
    \begin{subfigure}[t]{0.3\textwidth}
        \centering
        \includegraphics[width=\linewidth,height = \linewidth]{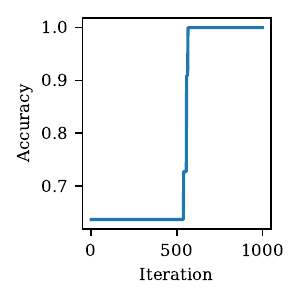} 
        \caption{Deepset} \label{fig:DSloss}
    \end{subfigure}
    \hfill
    \begin{subfigure}[t]{0.3\textwidth}
        \centering
        \includegraphics[width=\linewidth,height = \linewidth]{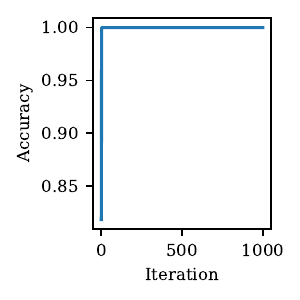} 
        \caption{Linear} \label{Lloss}
    \end{subfigure}
    \caption{Loss and accuracy of different models during training performed on validaiton dataset. \label{fig1}}
    \end{adjustbox}
\end{figure}

CloudPred had optimal performance with a mix of $10$ Gaussian distributions. Although during the training, all models reached an accuracy of $1$ over $1000$ iterations, CloudPred reached its optimal parameters corresponding to a loss and accuracy of $0$ and $1$, respectively, as early as the training started (\textbf{Fig. \ref{fig1}}). In comparison, the Deepset model required over 500 iterations to achieve training accuracy above 0.9 and the linear model minimized training losses and converged early in training.

CloudPred also identifies a single component that has the highest variability across patients and is sufficient for predicting patient states with high accuracy. The cell type composition of this component is listed in \textbf{Table \ref{tab2}}. The ability of CloudPred in providing interpretable results and biological insight on scRNA-seq analysis is highlighted in \textbf{Fig. \ref{fig:pca_tsne}}. \textbf{Fig. \ref{fig:pca_tsne},a} shows the projection of the test dataset on two first principal components derived during training and \textbf{Fig. \ref{fig:pca_tsne},b} illustrates a 2-component tSNE transformation of the data. Each ellipse represents a Gaussian component and the highlighted ellipse represents the cluster with highest prediction capability.

\begin{table}[!t]
\caption{Cellular composition of the cluster with highest prediction capability \label{tab2}}%
\begin{tabular*}{\columnwidth}{@{\extracolsep\fill}lcl@{\extracolsep\fill}}
\toprule
Cell type &  \begin{tabular}{@{}c@{}}Percent of contribution \\ in the most predictive cluster\end{tabular}  \\
\midrule
TAMs    & 0.35  \\
Plasma cells    & 0.24    \\
Epithelial    & 0.11    \\
CAFs    & 0.1    \\
Cycling T cells & 0.06    \\
Plasma T cells    & 0.03    \\
Cycling epithelial        & 0.02    \\
Endothelial  & 0.02    \\
Pericytes  & 0.01    \\
DCs  & 0.01    \\

\end{tabular*}

\end{table}

\section{Discussion}\label{sec3}
In this paper, we have demonstrated CloudPred's ability to correctly classify a breast cancer patient's disease status using scRNA-seq data from healthy patients and those with TNBC. We also compared its performance against a linear model and a Deepset model.

The CloudPred algorithm was able to correctly predict a patient's phenotype early during the training by identifying components with high variability using an unsupervised Gaussian mixture model. During training, the initialized mixture model performs so well that further training was not required to achieve an AUC and accuracy of 1. We reason that the breast cancer dataset used in this study has distinguished features that enables a simple Gaussian mixture model to successfully predict a healthy vs. diseased state. The breast cancer scRNA-seq dataset used in this study, however, had a limited number of samples of 21 patients. In comparison, the original study of He \textit{et al.} \cite{he2021arxiv} had access to scRNA-seq data of 142 patients containing the sequencing data of more than 500,000 cells. In larger datasets, the non-homogeneity of samples increases enough so that further refinement of the Gaussian mixture components will be required. However, we note that when we dived deeper, we were unable to find the raw patient data to reproduce the initial results in the original paper. The He \textit{et al.} \cite{he2021arxiv} study also found that CloudPred, Deepset, and the simple linear models all performed well for predicting lupus in patients and attributed the high accuracy of all three models to the prominent features of lupus. Incorporating additional breast cancer data and including different breast cancer cell lines (e.g. ER+, HER2+) may increase heterogeneity of scRNA-seq data \cite{Wang2018} and improve generalisation across patients and cancer types. 

A major advantage of CloudPred over the linear and Deepset models is that CloudPred provides direct intuitive evidence on what components were main drivers of the prediction task. In particular, CloudPred identifies one component that has highest variations across patients and is sufficient on its own for predicting patient state with highest accuracy. The cells that belong to this component are then characterized as predictors that vary the most among healthy and diseased patients. With a prior knowledge of cell types, one can utilize CloudPred's extracted predictors to identify cell types that play critical roles in patient's state determination. While He \textit{et al.} \cite{he2021arxiv} found the prevalence of CD4 T helper cells to be highly predictive of patient state, for our dataset, a mixture of different cell types were found to be highly prevalent in the component with best prediction accuracy ($AUC = 1$). 
The ability to identify relevant cell types within clusters may aid in the interpretability of results and can inform predictors of breast cancer disease states.
In this breast cancer dataset, the three of the major identified cell types were tumor-associated macrophages (TAMs), plasma cells, and epithelial cells, which accounted for 70\% of the most predictive cluster (Table \ref{tab2}). 

\begin{figure}
    \centering
    \includegraphics[scale = 0.85]{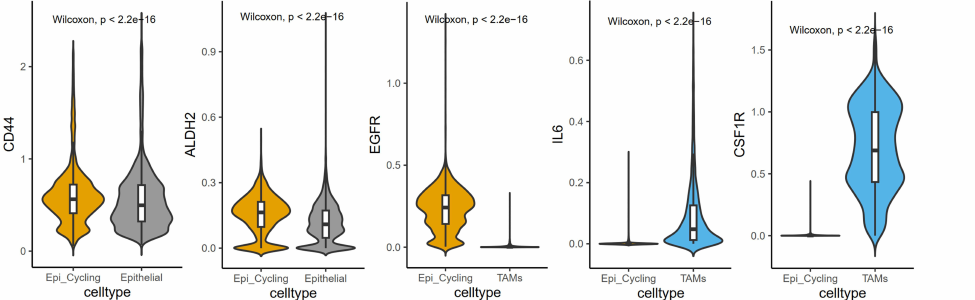}
    \caption{Expression of genes that help identify a crosstalk between BCSCs and TAMs. Genes from left to right: CD44, ALDH2, EGFR, IL6, CSF1R. Yellow, grey, and blue violins represent cycling epithelial cells, epithelial cells, and TAMs, respectively. (Wilcoxon rank test p-value $<$ 2.2e-16)}
    \label{fig:vlnplots}
\end{figure}

A prominent cancer marker is Ki67 coded by MKI67 gene. In Pal \textit{et al.}’s study~\cite{palSinglecellRNAExpression2021}, MKI67 expression was used to annotate cycling epithelial cells. MKI67 is expressed in 4 cell cycle phases (\textit{i.e.}, G1, S, G2, M) and is maximally expressed in mitosis. The single cell breast cancer atlas includes cycling epithelial cells in all histological subtypes of breast cancer (i.e., ER+, HER2+, and TNBC), but the TNBC has the highest ratio of cycling epithelial cells to total epithelial cells. It has reported that Ki67 knockout suppresses stem cell characteristics~\cite{mroujKi67RegulatesGlobal2021}. TNBC is distinguished by the largest population of breast cancer stem cells (BCSCs)~\cite{guhaCancerStemCell2023}. To investigate whether the cycling epithelial population contains the BCSCs, the expression of two BCSC markers (\textit{i.e.}, aldehyde dehydrogenase (ALDH2) or CD44) were compared between epithelial and cycling epithelial cells (Fig.~\ref{fig:vlnplots}). Significantly higher expression of ALDH2 and CD44 cycling epithelial cells suggests the cycling epithelial population may include some BCSCs. Ge \textit{et al.} described a crosstalk between TAMs and BCSCs in which TAMs exhibit multiple functions in tumor microenvironment that promote tumor progression~\cite{guhaCancerStemCell2023}. TAMs decrease immune surveillance by expressing immunosuppressive receptors and secrete cytokines that foster tumor proliferation~\cite{guhaCancerStemCell2023}.
TAMs secrete IL-6, IL-8, IL-10, and epidermal growth factor (EGF) which activate the STAT3 signaling pathway in BCSCs. BCSCs, in return, enhance the tumorigenicity of TAMs by secreting CSF-1.
We investigated this crosstalk in our data (Fig.~\ref{fig:vlnplots}) by comparing the expression of EGF receptor (EGFR), IL-6, IL-8, CSF-1, and CSF-1 receptor (CSF1R). Interestingly, we observed high expression of pro-inflammatory genes IL6, and CXCL8 (IL-8) in TAMs. The CSF1 and CSF1R expressions are higher in the cycling epithelial cells and TAMs, respectively.  
Additionally, the importance of TAMs for prediction between healthy vs. diseased states aligns with studies that show TAMs are the most abundant immune cells in the breast tumor microenvironment \cite{Huang2021} which is also coherent with our TNBC data (Fig.~\ref{fig:tsne_normal_tnbc}D). Epithelial cells have also been hypothesized to have a significant contribution to breast cancer \cite{Peng2020}.

Overall, we found that CloudPred can successfully predict a patient's breast cancer disease state using scRNA-seq data and can identify cell types with high variability that may be used as phenotypic predictors. CloudPred was able to achieve higher classification accuracy compared to common machine learning methods, such as linear and Deepset models. This study validates the performance of CloudPred on a different dataset and a disease compared to the original study \cite{he2021arxiv}, which extends the use of this method beyond lupus and race classification. 

\section{Conclusion}

CloudPred is an interpretable machine learning algorithm that can predict a patient's disease phenotype from their scRNA-seq data. In this study, we demonstrated CloudPred's ability to predict a patient's breast cancer disease status and to identify relevant cell subpopulations that may be used as predictors of disease phenotype. CloudPred was able to achieve an AUC of 1 and 100\% accuracy during validation and testing and its performance was compared to a linear model (95\% accuracy) and Deepset model (94\% accuracy). Future work should investigate its performance on larger datasets, as the selected dataset within this study only included a total 21 patients, as well as different diseases that may benefit from scRNA-seq based analysis.

\section{Acknowledgements}

T.C., B.M., H.M., and M.W. planned the proposal and wrote the manuscript. B.M. prepared and preprocessed the data. H.M. analyzed the data with CloudPred, linear, and Deepset models.

\bibliographystyle{abbrv}
\bibliography{paper.bbl}

\begin{thebibliography}{1}

\bibitem{chenCodeDownstreamAnalysis2022}
Y.~Chen, B.~Pal, G.~J. Lindeman, J.~E. Visvader, and G.~K. Smyth.
\newblock R code and downstream analysis objects for the {{scRNA-seq}} atlas of normal and tumorigenic human breast tissue.
\newblock {\em Scientific Data}, 9(1):96, Mar. 2022.

\bibitem{guhaCancerStemCell2023}
A.~Guha, K.~K. Goswami, J.~Sultana, N.~Ganguly, P.~R. Choudhury, M.~Chakravarti, A.~Bhuniya, A.~Sarkar, S.~Bera, S.~Dhar, J.~Das, T.~Das, R.~Baral, A.~Bose, and S.~Banerjee.
\newblock Cancer stem cell\textendash immune cell crosstalk in breast tumor microenvironment: A determinant of therapeutic facet.
\newblock {\em Frontiers in Immunology}, 14, 2023.

\bibitem{he2021arxiv}
B.~He, M.~Thomson, M.~Subramaniam, R.~Perez, C.~J. Ye, and J.~Zou.
\newblock Cloudpred: Predicting patient phenotypes from single-cell rna-seq.
\newblock {\em arXiv.org}, 2021.

\bibitem{Huang2021}
X.~Huang, J.~Cao, and X.~Zu.
\newblock Tumor-associated macrophages: An important player in breast cancer progression.
\newblock {\em Thorac Cancer}, 13(3):269--276, Dec. 2021.

\bibitem{mroujKi67RegulatesGlobal2021}
K.~Mrouj, N.~{Andr{\'e}s-S{\'a}nchez}, G.~Dubra, P.~Singh, M.~Sobecki, D.~Chahar, E.~Al~Ghoul, A.~B. Aznar, S.~Prieto, N.~Pirot, F.~Bernex, B.~Bordignon, C.~{Hassen-Khodja}, M.~Villalba, L.~Krasinska, and D.~Fisher.
\newblock Ki-67 regulates global gene expression and promotes sequential stages of carcinogenesis.
\newblock {\em Proceedings of the National Academy of Sciences of the United States of America}, 118(10):e2026507118, Mar. 2021.

\bibitem{palSinglecellRNAExpression2021}
B.~Pal, Y.~Chen, F.~Vaillant, B.~D. Capaldo, R.~Joyce, X.~Song, V.~L. Bryant, J.~S. Penington, L.~Di~Stefano, N.~Tubau~Ribera, S.~Wilcox, G.~B. Mann, {kConFab}, A.~T. Papenfuss, G.~J. Lindeman, G.~K. Smyth, and J.~E. Visvader.
\newblock A single-cell {{RNA}} expression atlas of normal, preneoplastic and tumorigenic states in the human breast.
\newblock {\em The EMBO Journal}, 40(11):e107333, June 2021.

\bibitem{Peng2020}
S.~Peng, L.~L. Hebert, J.~M. Eschbacher, and S.~Kim.
\newblock Single-cell rna sequencing of a postmenopausal normal breast tissue identifies multiple cell types that contribute to breast cancer.
\newblock {\em Cancers}, 12(12), 2020.

\bibitem{Wang2018}
F.~Wang, Z.~Dohogne, J.~Yang, Y.~Liu, and B.~Soibam.
\newblock Predictors of breast cancer cell types and their prognostic power in breast cancer patients.
\newblock {\em BMC Genomics}, 19(1):137, Feb 2018.

\end{thebibliography}

\end{document}